\documentclass[prb,twocolumn,groupedaddress,a4paper,showpacs,amssymb,amsmath]{revtex4}
\usepackage{graphicx}

\begin{document}

\title{Phase diffusion and fractional Shapiro steps in 
superconducting quantum point contacts}

\author{Rapha\"el Duprat}
\author{Alfredo Levy Yeyati}

\affiliation{Departamento de F\'isica Te\'orica de la Materia Condensada, Facultad de Ciencias C-V, Universidad Aut\'onoma de Madrid, E-28049 Madrid, Spain}

\date{\today}

\begin{abstract}
We study the influence of classical phase diffusion 
on the fractional Shapiro steps in resistively shunted
superconducting quantum point contacts. The problem is mapped onto a  
Smoluchowski equation with a time dependent potential. A numerical 
solution 
for the probability density of the phase difference between the leads 
gives 
access to the mean current and the mean voltage across the contact. 
Analytical 
solutions are derived in some limiting cases. We find that the effect of 
temperature is stronger on fractional than on integer steps, in accordance 
with preliminary experimental findings. We further extend the analysis to a 
more general environment including two resistances and a 
finite capacitance.
\end{abstract}


\maketitle
\section{Introduction}
\label{intro}
Shapiro steps\cite{shapiro63} in superconducting tunnel junctions are a clear 
evidence of the a.c. Josephson effect\cite{josephson62}, {\it i.e.} the flow 
of an alternating current through the junction under a finite dc voltage. 
Shapiro steps appear as a consequence of the beating between an applied 
microwave field and this alternating current. Their observation thus provides 
an indirect test of the sinusoidal current-phase relation. For the case of 
highly transmissive junctions, like atomic size contacts, the presence of 
higher harmonics in the current-phase relation leads to fractional Shapiro 
steps \cite{cuevas02}. 

In contrast to conventional tunnel junctions, atomic contacts are
characterized by a reduced set of conduction channels whose
transmissions $\tau_n$ can take arbitrary values between 0 and 1
\cite{review03}. The channel content of a given contact can be determined 
experimentally with high accuracy by analyzing the subgap structure of
the $IV$ characteristic in the superconducting state
\cite{scheer97}. Moreover, the electromagnetic environment of
the contact can be designed by means of litographic techniques
\cite{goffman00}. Due to all these properties atomic
contacts can be considered as ideal systems to test theoretical
predictions on mesoscopic electron transport under controlled
conditions \cite{review03}.       

Shapiro steps in a superconducting
atomic contact of arbitrary transmission
were analyzed in Ref. \onlinecite{cuevas02} 
within a fully microscopic approach in which an
ideal voltage bias on the contact was assumed. 
However, a more realistic description of this phenomenon 
requires to take into account 
environmental effects, the most important being certainly the effect 
of phase diffusion caused by thermal noise in the circuit in which the
contact is embedded.

The influence of thermal fluctuations on the current-voltage characteristics 
in tunnel junctions has been traditionally analyzed by means of the
so-called resistively and capacitively shunted junction (RCSJ) model
\cite{tinkham}.
The starting point of this approach is to write down a Langevin equation for 
the phase difference accross the junction \cite{ivan69,ambegaokar69}. 
In this article, we
generalize this model to the case of highly transmissive junctions
in the presence of a microwave field in order to study the 
effect of temperature on fractional Shapiro steps in atomic size contacts. 
Our work is also motivated by experiments underway in the Quantronics 
group at the C.E.A. in Saclay. Their preliminary results seem to  
indicate that fractional steps are more affected by thermal fluctuations 
than the integer ones \cite{saclay}.

This study will be based on two main assumptions: 1) the external frequencies
and thus the range of voltages considered are small compared to the
superconducting gap on the contact leads. This would allow us to assume an
{\it adiabatic} response of the contact following its static current-phase
relation and 2) the resistances in the circuit containing the contact are
small compared to the resistance quantum so as to neglect quantum effects 
in our treatment. As we discuss below, low values of the shunting resistances
are also necessary in order to observe well defined 
Shapiro steps even when thermal fluctuations are negligible.

The paper is organized as follows: In the first section we present our 
generalization of the RCSJ model for contacts of arbitrary transmission
in the presence of  microwave radiation. 
The second section describes how to solve the 
stochastic equation of the problem and shows the numerical results for the 
current-voltage characteristics. In the third section we discuss some 
limiting cases in which analytical results can be obtained.
In the last section we consider a more general electromagnetic
environment including two resistances and a finite capacitance and
discuss its effect on the supercurrent peak in the absence of radiation.
We finally present some concluding remarks.

\section{\label{sec:model}The model}

\subsection{Electrical circuit}

\begin{figure}[h!]
\includegraphics[width=\columnwidth]{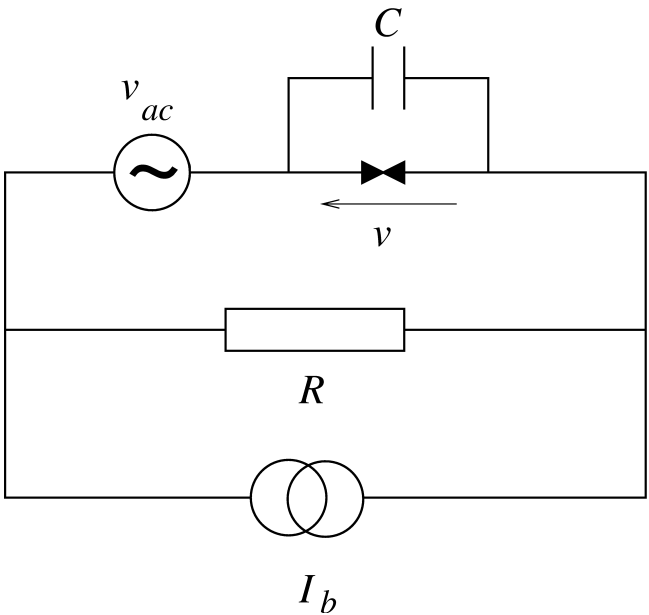}
\caption{\label{fig:circuit} Equivalent circuit of the resistively and 
capacitively 
shunted junction (RCSJ) model in presence of a microwave field
$v_{ac}\cos\omega t$.}
\end{figure}

We consider the standard RCSJ model, to which we add a microwave field. 
The equivalent circuit is shown on the figure \ref{fig:circuit}. The 
parallel combination of the contact with a resistance $R$ and a 
capacitance $C$ is current polarized by a current source $I_b$.
The microwave field introduces an additional ac voltage, $v_{ac}$ 
on the branch containing the contact.
We denote by $\theta$ 
the superconducting phase difference across the contact and by
$I(\theta)$ the corresponding current-phase relation.
Dissipation in the 
resistance will cause Johnson-Nyquist noise $L(t)$. Conservation of 
the current implies that
\begin{equation} \label{cc}
I_b = C\frac{dv}{dt} + I(\theta) + \frac{V}{R} + L(t) \ ,
\end{equation}
with $V=v+v_{ac}\cos\omega t$.
The voltage across the contact is related to the phase via the usual relation 
$v=\phi_0 \dot{\theta}$ ($\phi_0 \equiv \hbar/2e$ being the reduced flux 
quantum). 

In order to stress the analogy with the Brownian motion of a particle in a
time-dependent potential, we follow Ref. \onlinecite{ambegaokar69}
and define $M\equiv\phi_0^2C$, $p\equiv\phi_0 vC$, 
$\eta\equiv 1/RC$, $r(t)\equiv -\phi_0 L(t)/\eta M$ and
\begin{equation}
U(\theta,t) \equiv \phi_0\theta \Bigl[I_b -\frac{v_{ac}}{R}\cos\omega t\Bigr] -\phi_0\int_0^\theta I(\theta) d\theta \ .
\end{equation}
The originality of this work lies in the explicit time dependence of the 
potential $U(\theta,t)$ and the non-trivial form of the current-phase 
relation $I(\theta)$ which we describe in more detail below. When the 
capacitance is negligible, we are in the strong damping regime 
$\eta p \gg \dot{p}$, and (\ref{cc}) takes the form of a simple Langevin 
equation
\begin{equation} \label{overdmpLan}
\frac{p}{M} = \frac{1}{\eta M} \frac{\partial U}{\partial \theta} + r(t) \ .
\end{equation}
We consider a gaussian white noise, that is
\begin{eqnarray} \label{rnoise}
<r(t)> &=& 0 \\
<r(t)r(t')> &=& \delta(t'-t) \frac{2T}{\eta M} \\
{\cal P}[r(t)] &=& \exp \biggl[-\frac{\eta M}{4T}\int r^2(t) dt \biggr]  \ ,
\end{eqnarray}
where we have set the Boltzmann constant equal to one.
From the overdamped Langevin equation (\ref{overdmpLan}), one can derive a Smoluchowski equation for the probability density of the phase $\sigma(\theta,t)$ (see Ref. \onlinecite{schwabl} for instance):
\begin{equation} \label{smolu}
\frac{d\sigma}{dt} = \frac{1}{\eta M} \frac{\partial}{\partial \theta} \biggl[-\frac{\partial U}{\partial \theta} \sigma + T\frac{\partial \sigma}{\partial \theta} \biggr] \ .
\end{equation}
Notice that the coefficients in this equation are not dependent on the 
contact capacitance, as expected for the strong damping regime. 
Floquet's theorem then tells us that its general solution has the form
\begin{equation}
\sigma(\theta,t) = e^{-\lambda t}\tilde{\sigma}(\theta,t)
\end{equation}
The exponential factor represents the relaxation towards the stationary 
solution $\tilde{\sigma}$ with a time constant $\lambda^{-1}$. 
Note that in our case, owing to the time dependence of 
the potential $U$, the stationary solution oscillates in time with
the external frequency $\omega$. 
The characteristic frequency given by $\omega_c = RI_c/\phi_0$, where 
$I_c = \mbox{max}[I(\theta)]$ is the critical current,
can be considered as the typical relaxation rate of the system
in the absence of thermal fluctuations \cite{vanneste}. 
As it is well known for current biased tunnel junctions 
\cite{vanneste}, the observation of well  
defined Shapiro steps in the $IV$ curves requires $\omega > \omega_c$.
Notice on the other hand that for the validity of the adiabatic 
approximation discussed in Sect. \ref{sec:model}-B 
we are requiring $\omega \ll \Delta_A$.
In the rest of this work we will only consider 
stationary solutions ($\lambda =0$) and will thus identify $\sigma$ with 
$\tilde{\sigma}$.

\subsection{Current-phase relation}

\begin{figure}
\includegraphics[width=\columnwidth]{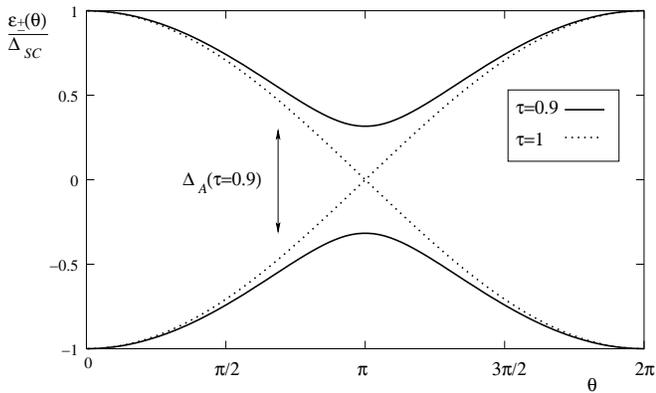}
\caption{\label{fig:andreev} Andreev levels as a function of the phase difference for $\tau=0.9$ and $\tau=1$.}
\end{figure}

A crucial ingredient of the model is the current-phase relation
$I(\theta)$. For simplicity we consider a contact with one conduction 
channel of transmission 
$\tau\in [0,1]$ (extension of the theory to the multichannel case is
straightforward) . As it is well known from the mesoscopic theory of the
Josephson effect \cite{beenakker92} the current through the contact is 
carried by the so-called Andreev states with energies given by
\begin{equation}
\epsilon_\pm(\theta) = \pm\Delta_{SC}\sqrt{1-\tau\sin^2\frac{\theta}{2}} \ ,
\end{equation}
($\Delta_{SC}$ being the superconducting gap) whose separation at $\theta = \pi$ is $\Delta_A = 2\Delta_{SC}\sqrt{1-\tau}$. Note that this Andreev gap closes in the ballistic limit $\tau \rightarrow 1$ (figure \ref{fig:andreev}). When the voltage is much smaller than the Andreev gap ($eV \ll \Delta_A$), one can assume the system to remain in the state of lowest energy (this is the {\it adiabatic approximation}) and obtain the following current-phase relation
\cite{beenakker92}:
\begin{equation} \label{current-phase-A}
I(\theta) = \frac{\Delta_{SC}}{4\phi_0} \frac{\tau\sin\theta}{\sqrt{1-\tau\sin^2\frac{\theta}{2}}} \tanh\biggl( \frac{\Delta_{SC}}{2T}\sqrt{1-\tau\sin^2\frac{\theta}{2}}  \biggr) \ .
\end{equation}

\begin{figure}
\includegraphics[width=\columnwidth]{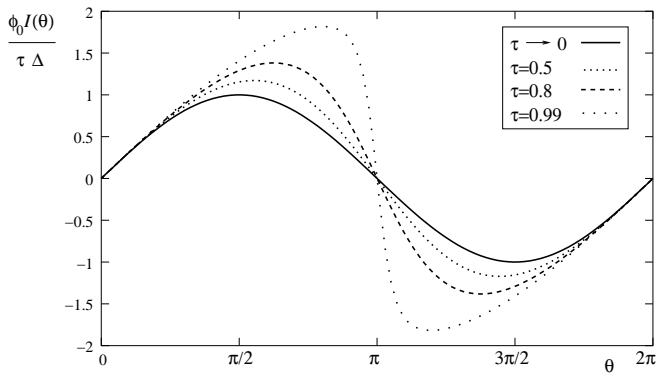}
\caption{\label{fig:AdiaCP} Evolution of the adiabatic current-phase relation (\ref{current-phase-A}) with the transmission $\tau$ at zero temperature.}
\end{figure}

When $\tau$ is small, one recovers the usual sinusoidal relation for
tunnel junctions with an expression of the critical current that matches
Ambegaokar-Baratoff formula\cite{ambegaokar63}. On the other hand, when
the contact is strongly transmissive the contribution of higher harmonics 
can no longer be neglected and the current-phase gradually approaches the
$\sin{(\theta/2)}$ behavior characteristic of the ballistic limit 
(figure \ref{fig:AdiaCP}). These higher harmonics give rise to 
fractional Shapiro steps when a microwave field is applied.

An estimate of the size of these steps at zero temperature can be obtained 
assuming that the total bias voltage $V$ remains constant.
The phase evolution is then given by $\theta(t) = Vt/\phi_0
-  v_{ac} \sin{(\omega t)}/\phi_0 \omega + \theta_0$.
In this limit and 
within the adiabatic
approximation one can thus obtain an expression for the size of the fractional
Shapiro steps by introducing this phase evolution into the current-phase
relation (\ref{current-phase-A}) and performing its Fourier decomposition. 
The step at $V=\phi_0 \omega n/k$ is then given by \cite{carvello02}

\begin{equation}
I_{\frac{n}{k}} = \sum_{m=1}^{\infty} I_{m \times k} J_{m \times n}(2mk\alpha)
\sin{m k \theta_0} (-1)^{m n},
\end{equation}
where $I_m$ denotes the harmonics in the current-phase relation,
$J_n$ are the integer order Bessel functions and 
$\alpha = ev_{ac}/\hbar\omega$.

When the voltage is comparable to the Andreev gap, the adiabatic approximation 
breaks down since one can no longer assume the system to stay in the lowest 
Andreev level. In the present work we shall restric our analysis to the 
adiabatic approximation and concentrate on the effects of phase
fluctuations in the current-voltage characteristics.

\section{Current-voltage characteristics}

This section is divided into three parts. In the first one we derive the 
expression of the mean current and the mean voltage in terms of the solution 
of the Smoluchowski equation. In the second one 
we derive an expression for this solution in terms of a matrix continued 
fraction. In the third part we show the numerical results for the $I-V$ 
characteristics.

\subsection{Expressions of the mean current and the mean voltage}

If we define $w \equiv  -\frac{1}{\eta M} \bigl[-\frac{\partial U}{\partial 
\theta} \sigma + T\frac{\partial \sigma}{\partial \theta} \bigr]$, we can 
rewrite (\ref{smolu}) as:
\begin{equation} \label{cont}
\frac{\partial \sigma}{\partial t} + \frac{\partial w}{\partial \theta} = 0
\end{equation}
This equation can be seen as a conservation law for the probability 
(which always holds for stationary solutions). Henceforth, $w$ is a 
probability current and must be given by:
\begin{equation} \label{probcurr}
w = \sigma \frac{d\theta}{dt} = \sigma \frac{v}{\phi_0}
\end{equation}

\begin{figure}
\includegraphics[width=\columnwidth]{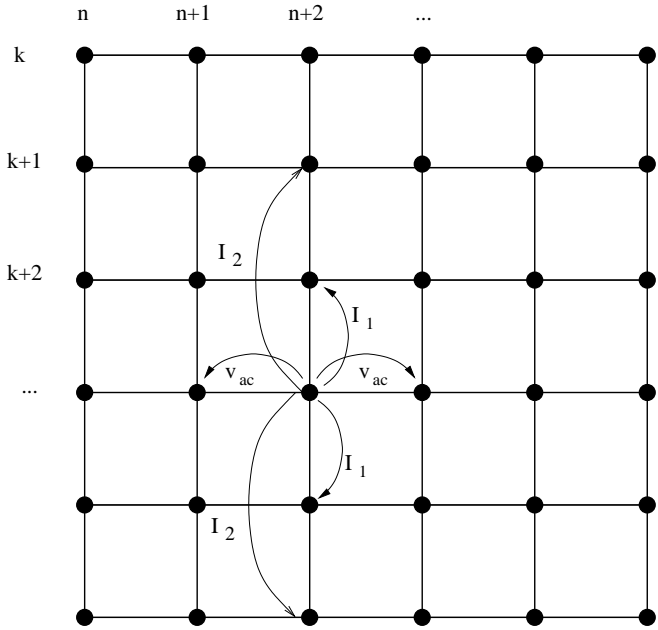}
\caption{\label{fig:lattice} Lattice representation of the Fourier-space Smoluchowski equation. The amplitude of the microwave $v_{ac}$ couples the $n$-chains, while the harmonics of the current-phase relation couple the $k$-chains.}
\end{figure}

Let us denote by $\overline{(...)}$ the mean value of a quantity 
with respect to the phase 
and $<(...)>$ its mean value with respect to time. Our aim is to calculate 
the current-voltage characteristics, that is $<\overline{I(\theta)}>$ as a 
function of $<\overline{v}>$, where
\begin{eqnarray}
<\overline{I(\theta)}> &=& \int_0^{2\pi} d\theta \int_{-\infty}^{\infty} 
dt \sigma(\theta,t) I(\theta) \\
<\overline{v}> &=& \phi_0 \int_0^{2\pi} d\theta \int_{-\infty}^{\infty} 
dt w(\theta,t) \ .
\end{eqnarray}

The Smoluchowki equation (\ref{smolu}) is periodic in time and phase, and 
so must be the density $\sigma(\theta,t)$ and the probability current 
$w(\theta,t)$. They can thus be expanded in double Fourier series:
\begin{eqnarray}
\sigma(\theta,t) &=& \sum_{n,k\in\mathbb{Z}} \sigma_{n,k}e^{ik\theta + in\omega t} \\
w(\theta,t) &=&  \sum_{n,k\in\mathbb{Z}} w_{n,k}e^{ik\theta + in\omega t} \ .
\end{eqnarray}

The normalization condition for the probability density then imposes that
\begin{equation} \label{normalisation}
\sigma_{n,0} = \delta_{n,0}/2\pi \ .
\end{equation}
In Fourier space, the Smoluchowski equation (\ref{smolu}) reads
\begin{widetext}
\begin{eqnarray} \label{Fouriersmolu}
in\omega \sigma_{n,k} &=& \frac{1}{\eta M} \biggl[ -k^2T\sigma_{n,k} 
-ik\phi_0 I_b\sigma_{n,k} +ik\phi_0\sum_m I_m (\sigma_{n,k-m} - \sigma_{n,k+m}) 
\biggr. \nonumber \\
& & \biggl. +k\phi_0\frac{v_{ac}}{2R}(\sigma_{n+1,k} + \sigma_{n-1,k}) \biggr] \ .
\end{eqnarray}
\end{widetext}

It is worth noticing that this set of equations can be associated with  
a (non-hermitian) lattice model for particles in a square lattice. 
Each component of the density $\sigma_{n,k}$ can
be associated with a site $(n,k)$  in a  
$\mathbb{Z}\times\mathbb{Z}^*$ square lattice. The coupling between
chains $n$ and $n\pm 1$ is proportinal to the ac-voltage $v_{ac}$, while the 
coupling between chains $k$ and $k\pm m$ is proportional to the $m$-th 
harmonic of the Josephson current $I(\theta)$ (see figure \ref{fig:lattice}). 
This analogy will be useful for deriving approximate analytical
solutions as discussed in Sect. \ref{sec:supercurrent}.

Note, on the other hand, that temperature appears 
in Eq. (\ref{Fouriersmolu}) with a factor $k^2$. 
We will see later that in the limit of large ac voltage 
$k^2T$ actually plays the role of an effective temperature for Shapiro steps 
of order $n/k$.

Making use of the orthogonality of circular functions, one 
can easily show that
\begin{eqnarray}
&<&\bar{v}> = 2\pi\phi_0 w_{0,0} = R \Bigl(I_b - \sum_{k\in\mathbb{Z}} 
\sigma_{0,k}I_{-k} \Bigr) \label{V-I} \\
&<&\overline{I(\theta)}> = \sum_{k\in\mathbb{Z}} 
\sigma_{0,k}I_{-k} \label{current} \ .
\end{eqnarray}

Thus, in order to calculate the current-voltage characteristics, we only 
need to know the $\sigma_{0,k}$'s. A Shapiro step of order $n/k$ in 
the $IV$ characteristics is precisely due to the presence of a jump in 
$\sigma_{0,k}$ as a function of the bias current.

\subsection{Recursive solution of the Smoluchowski equation}

We now expose how to solve (\ref{Fouriersmolu}) numerically in order to 
obtain the Fourier components of the probability density.

It is convenient to introduce the vectors 
$\vec{\sigma}_n$ $\equiv$ 
$(...,\sigma_{n,2},\sigma_{n,1},\sigma_{n,-1},\sigma_{n,-2},...)$, 
$\vec{I}$ $\equiv$ $(...,I_{-2},I_{-1},I_1,I_2,...)$ and the matrices 
$\mathbf{A}_n$ defined by :
\begin{eqnarray}
(A_n)_{kk'} \equiv & &\Bigl(\frac{n\omega \eta M}{k} -ikT +\phi_0 I_b\Bigr)\delta_{kk'} \nonumber\\
& & -I_m\phi_0 (\delta_{k',k-m} - \delta_{k',k+m}) \ .
\end{eqnarray}
We can then rewrite (\ref{Fouriersmolu}) in a more compact form,
\begin{equation}
\mathbf{A}_n \vec{\sigma}_n = \phi_0\frac{v_{ac}}{2R}(\vec{\sigma}_{n+1}+
\vec{\sigma}_{n-1}) + \delta_{n,0}\frac{\phi_0}{2\pi}\vec{I} \ .
\end{equation}
From now on, we take $n>0$, and define $\underline{n} \equiv -n$. If we define the matrices $\mathbf{S}_n$ and $\mathbf{S}_{\underline{n}}$ by
\begin{eqnarray}
\mathbf{S}_{n+1} \vec{\sigma}_n = \phi_0\frac{v_{ac}}{2R}\vec{\sigma}_{n+1} \\
\mathbf{S}_{\underline{n+1}} \vec{\sigma}_{\underline{n}} = \phi_0\frac{v_{ac}}{2R}\vec{\sigma}_{\underline{n+1}} \ ,
\end{eqnarray}
we obtain that
\begin{equation} \label{MCF1}
\vec{\sigma}_0 = \Bigl[ \mathbf{A}_0 -\mathbf{S}_1 -\mathbf{S}_{\underline{1}}  \Bigr]^{-1} \frac{\phi_0}{2\pi}\vec{I} \ ,
\end{equation}
with
\begin{widetext}
\begin{equation} 
\mathbf{S}_{1(\underline{1})} = - \Bigl( \frac{v_{ac}}{2R}\phi_0 \Bigr)^2\cfrac{1}{\mathbf{A}_{1(\underline{1})}-\Bigl( \frac{v_{ac}}{2R}\phi_0 \Bigr)^2\cfrac{1}{\mathbf{A}_{2(\underline{2})}-\Bigl( \frac{v_{ac}}{2R}\phi_0 \Bigr)^2\cfrac{1}{\mathbf{A}_{3(\underline{3})} -...}}} \ .
\end{equation}
\end{widetext}

A recursive numerical solution of this last equation enables us to find $\vec{\sigma}_0$, and thus the mean voltage $<\bar{v}>$ and the mean current $<\overline{I(\theta)}>$. The accuracy of the results depends on the number of harmonics considered (in both, phase and time), that is on the size of the matrices 
($k_{max}$) and the cut-off in the continued fraction ($n_{max}$). 
As the temperature is lowered and the ac voltage is increased
the values of $n_{max}$ and $k_{max}$
required to get a good precision increase. 
The numerical results exposed below were obtained with 
$n_{max}=100$ and $k_{max}=50$ which where found to be sufficient 
to get reliable results at the lower temperatures considered.

\subsection{Numerical results}

\begin{figure}
\includegraphics[width=\columnwidth]{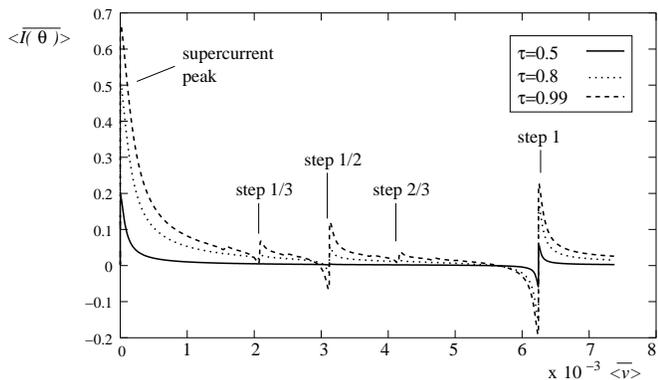}
\caption{\label{fig:IVdeTau} Mean current as a function of the mean voltage 
across the contact, for three different values of the transmission and the 
following parameters (expressed in reduced units as indicated in the text): 
$R=10^{-4}$, $T=10^{-2}$, $v_{ac}=4.4\times10^{-3}$, 
$\omega/2\pi = 12.5\times10^{-3}$.}
\end{figure}
\begin{figure}
\includegraphics[width=\columnwidth]{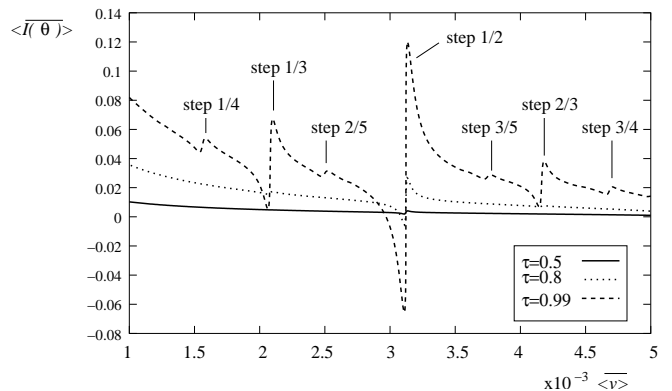}
\caption{\label{fig:IVdeTauZoom} Zoom of FIG. \ref{fig:IVdeTau}.}
\end{figure}

In order to illustrate the type of results that are obtained by the numerical
solution of Eq. (\ref{Fouriersmolu}) we have chosen parameters which roughly 
correspond to a possible experiment based on Al atomic contacts \cite{saclay}. 
The energy scale in such an experiment is set by the superconducting 
gap $\Delta_{SC} \simeq 180 \mu eV$. Typical microwave frequencies used
in experiments are  $\hbar \omega \sim 10^{-2}-10^{-1} \Delta_{SC}$.
On the other hand, as we stated in the introduction, the series 
resistance should be small compared to the resistance quantum  
$R_Q=h/4e^2$ in order to neglect  
Coulomb blockade effects at small temperatures \cite{nazarov}.  
Moreover, the observability of the fractional steps requires that
$\omega > RI_c/\phi_0$, i.e. much larger than the typical relaxation
rate in Eq. (\ref{smolu}). As $I_c \sim \Delta_{SC}/\phi_0$, this condition
implies that $R < 10^{-2} R_Q$.  
Figure \ref{fig:IVdeTau} shows the current-voltage characteristics for
different values of the transmission for a set of parameters chosen
according to this criterium.
Resistance is measured in units of $R_Q$ 
and all energies in units of the superconducting gap
$\Delta_{SC}$ ({\it e.g.} in units of $\Delta_{SC}/e$ for the voltage and 
$\Delta_{SC}/\phi_0$ for the current).
Naturally, the overall current
increases with the transmission $\tau$. As expected, additional Shapiro
steps appear as $\tau$ raises. Close to perfect transmission
($\tau=0.99$), one can clearly distinguish steps $1$, $1/2$, $1/3$,
$2/3$, $1/4$, $3/4$, $2/5$ and $3/5$ (figure \ref{fig:IVdeTauZoom}).

\begin{figure}
\includegraphics[width=\columnwidth]{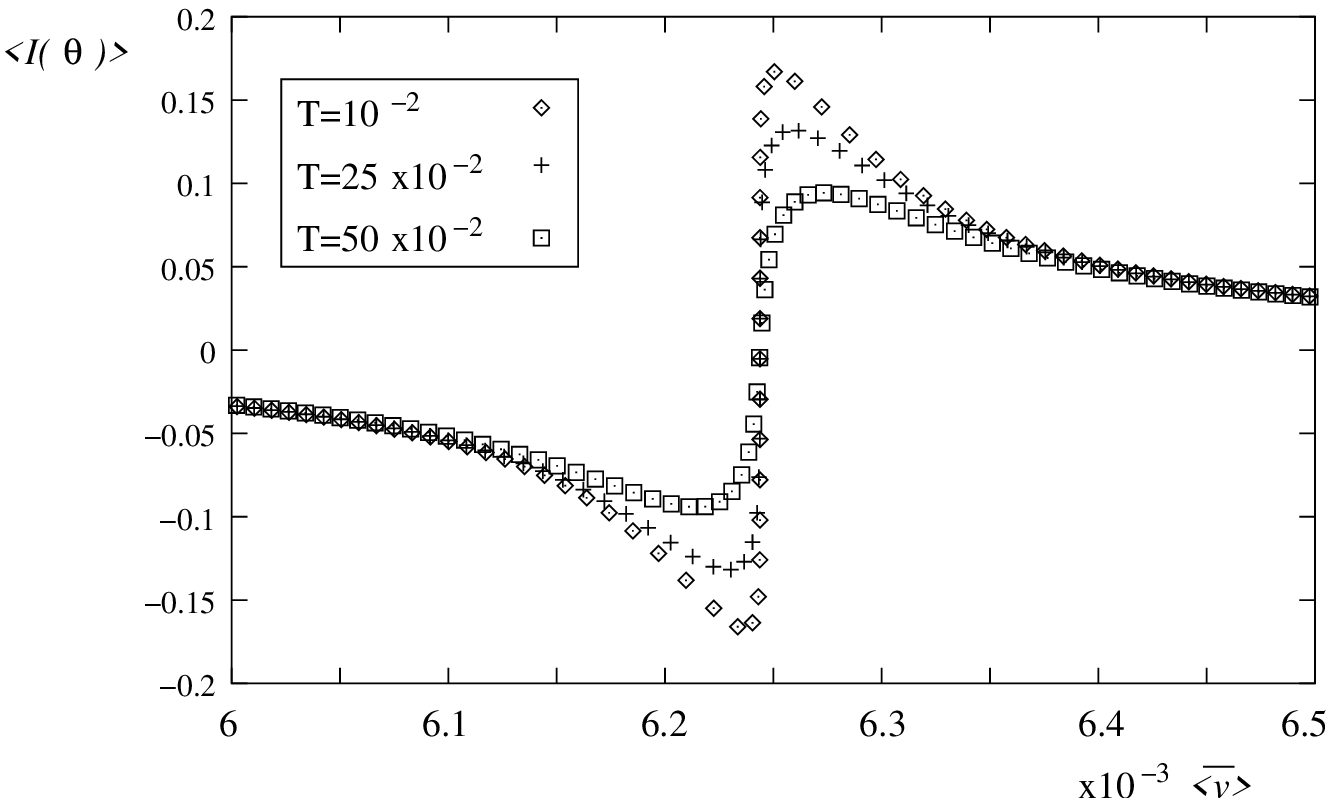}
\caption{\label{fig:pas1} Evolution of step 1 with temperature.}

\vspace{1cm}

\includegraphics[width=\columnwidth]{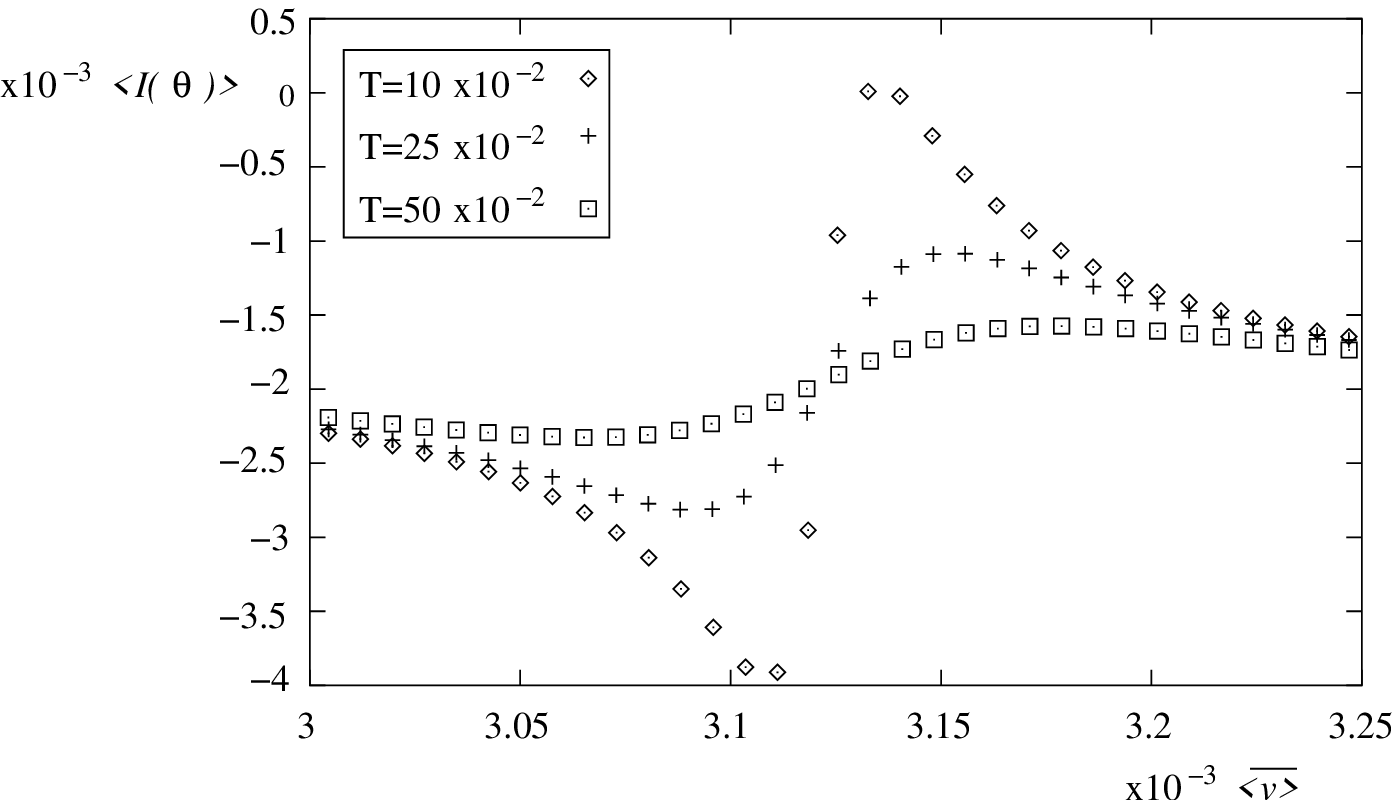}
\caption{\label{fig:pas1demi} Evolution of step 1/2 with temperature. The decrease of the amplitude is clearly much stronger than for step 1.}
\end{figure}

The temperature dependence of the I-V characteristic is illustrated in  
Figs. \ref{fig:pas1} and \ref{fig:pas1demi} showing the behavior of 
steps 1 and 1/2 respectively. As already anticipated, a stronger suppression
is observed for the 1/2 step. We shall analyze the scaling of the fractional 
steps with temperature in more detail in the next section. 
 
\section{\label{sec:supercurrent}Approximate analytical solutions}

In the quest for analytical expressions for $\sigma(\theta)$
it is useful to have in mind the lattice representation of Eq.
(\ref{Fouriersmolu}). We shall consider two limiting cases corresponding to
situations in which the {\it vertical} (i.e. labelled by $n$) or 
{\it horizontal} (i.e. labelled by $k$) chains are nearly decoupled, 
that is when 
the ac-voltage (resp., the Josephson current) is small. 
Notice that for the typical
choice of parameters discussed in the previous section $R I(\theta) \ll v_{ac}$, which
corresponds to this last case.
The spirit of this approach is to 
decouple the $n$ and $k$ dependences of the Fourier components 
of the probability density. 
This decoupling will then lead to recurrence relations between the 
harmonics that 
ressemble to those of Bessel and modified Bessel functions:
\begin{eqnarray}
\frac{2\nu}{z}{\cal{J}}_\nu(z) &=& {\cal{J}}_{\nu-1}(z) + {\cal{J}}_{\nu+1}(z) \\
\frac{2\nu}{z}{\cal{I}}_\nu(z) &=& {\cal{I}}_{\nu-1}(z) - {\cal{I}}_{\nu+1}(z) \ .
\end{eqnarray}
Like in Ref. \onlinecite{ivan69}, this analogy can be used to obtain analytical solutions within this weak interchain coupling approximation.

\subsection{Limit of small ac-voltage and low transmission}

Let us define $\{\lambda_k ; k\in\mathbb{Z}^*\}$ as the solution of the Smoluchowski equation (\ref{Fouriersmolu}) in the absence of microwaves 
($v_{ac} = \omega = 0$). In the tunnel limit $I(\theta)=I_c\sin\theta$, 
it obeys the equation
\begin{equation}
(-ikT+\phi_0 I_b)\lambda_k = \phi_0\frac{I_c}{2i}(\lambda_{k-1}-\lambda_{k+1}) \ .
\end{equation}
Ivanchenko and Zil'berman\cite{ivan69} found its solution by noting the analogy with the recurrence relation for modified Bessel functions ${\cal{I}}_\nu(z)$:
\begin{equation}
\lambda_k = \Theta(-k) \frac{{\cal{I}}_{k+i\frac{\phi_0I_b}{T}}\Bigl(\frac{\phi_0I_c}{T}\Bigr)}{{\cal{I}}_{i\frac{\phi_0I_b}{T}}\Bigl(\frac{\phi_0I_c}{T}\Bigr)} + \Theta(k) \frac{{\cal{I}}_{k-i\frac{\phi_0I_b}{T}}\Bigl(\frac{\phi_0I_c}{T}\Bigr)}{{\cal{I}}_{-i\frac{\phi_0I_b}{T}}\Bigl(\frac{\phi_0I_c}{T}\Bigr)} \ .
\end{equation}
To arrive to this expression, one has to impose that the probability density
is real and normalized ($\Theta$ represents the Heavyside function).
If we now introduce a weak coupling $v_{ac}$ between the $n$-chains, we can try the following ansatz for the Fourier components: $\sigma_{n,k}=\kappa_{n,k}\lambda_k$, with $\kappa_{n,k}\simeq\kappa_{n,k\pm1}$. The $\kappa_{n,k}$'s are then solution of
\begin{equation}
\frac{2n}{k}\frac{\omega\phi_0}{v_{ac}} \kappa_{n,k} = \kappa_{n-1,k}+\kappa_{n+1,k} \ .
\end{equation}
Making use of the analogy with the recurrence relation of Bessel functions ${\cal{J}}_\nu(z)$, we obtain
\begin{equation}
\kappa_{n,k} \propto {\cal{J}}_n(2k\alpha) \ .
\end{equation}
We thus obtain a generalized expression of the Ivanchenko-Zil'berman solution for the normalized probability density in the limit of small $\alpha$ :
\begin{eqnarray}
\sigma_{n,k} =& & \frac{{\cal{J}}_n(2k\alpha)}{2\pi} \Biggl[ \Theta(-k) \frac{{\cal{I}}_{k+i\frac{\phi_0I_b}{T}}\Bigl(\frac{\phi_0I_c}{T}\Bigr)}{{\cal{I}}_{i\frac{\phi_0I_b}{T}}\Bigl(\frac{\phi_0I_c}{T}\Bigr)} \nonumber\\
&+& \Theta(k) \frac{{\cal{I}}_{k-i\frac{\phi_0I_b}{T}}\Bigl(\frac{\phi_0I_c}{T}\Bigr)}{{\cal{I}}_{-i\frac{\phi_0I_b}{T}}\Bigl(\frac{\phi_0I_c}{T}\Bigr)}  \Biggr] \ .
\end{eqnarray}


\subsection{Limit of small Josephson current}

\begin{figure}
\includegraphics[width=\columnwidth]{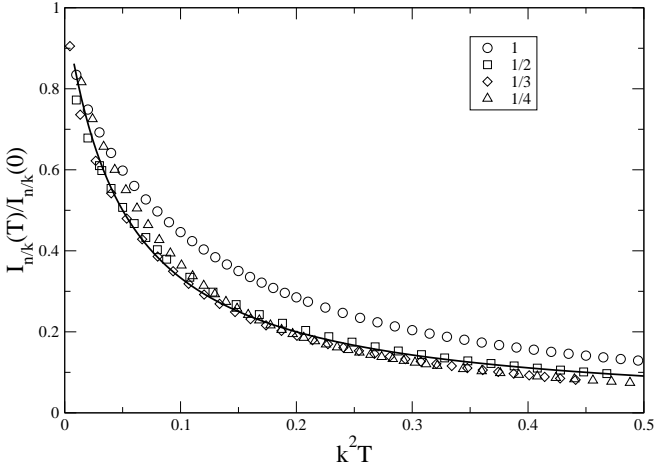}
\caption{\label{fig:pick2T} Scaling of the size of Shapiro steps 
of order $n/k$ with temperature for a contact with $\tau=0.99$
at fixed $k\alpha = 2.5$ and $R=10^{-3}$. The full line corresponds 
to a fit of the form $1/(1+ak^2T)$.}
\end{figure}

In this limit the coupling between $k$ chains is negligible. To zero
order in the Josephson current the Fourier components of the   
probability density satisfy the equation
\begin{eqnarray}
\Bigl(\frac{n\omega \eta M}{k} -ikT +I_b\phi_0\Bigr)\sigma^{(0)}_{n,k} 
= \frac{v_{ac}}{2R}\phi_0 \left( \sigma^{(0)}_{n+1,k} +
\sigma^{(0)}_{n-1,k} \right) \ .
\end{eqnarray}

Hence $\sigma^{(0)}_{n,k}$ is simply given by $J_{n+\nu_k}(2 k \alpha)/2\pi$, 
where $\nu_k = (-ik^2RT+ kRI_b\phi_0)/(\omega\phi_0^2)$. 
From (\ref{V-I}) and (\ref{current}) we see that, in order to calculate 
the $I-V$ curves we 
only need $\sigma^{(0)}_{0,k}$ and that it is the only quantity involving 
temperature. Thus, the mean current within this approximation is given
by 

\begin{eqnarray}
<\overline{I(\theta)}> \approx \frac{1}{2\pi} \sum_k J_{\nu_k}(2 k\alpha) I_{-k}.
\label{approx-IV}
\end{eqnarray}

Within this approximation a Shapiro step of order $n/k$ 
arise from the terms in $I_{\pm k}$ when
$RI_b \approx n\omega\phi_0/k$.
The above expression suggests that the size of the step should scale as
$J_{\nu_k}(2k\alpha)$, i.e. a universal temperature behaviour should be
observed when the size of the step is plotted against $k^2T$ for 
fixed $k\alpha$. 

This approximate scaling is illustrated in Fig. \ref{fig:pick2T}. 
As can be observed, the exact numerical results in this limit 
fulfil reasonably this predicted universal behavior.
At large temperatures
this can be well approximated by a universal function of the form
$1/(1+ak^2T)$, shown as a full line in Fig. \ref{fig:pick2T}.
However, it should be noticed that some deviation from universality
is always observed due to the finite coupling between the $k$-chains
not included in the approximation given by Eq. (\ref{approx-IV}).  
This deviation is more pronounced for the first Shapiro step ($k=1$)
which arises from the larger component of the current phase relation.

\section{A more realistic circuit}

\begin{figure}[ht]
\includegraphics[width=\columnwidth]{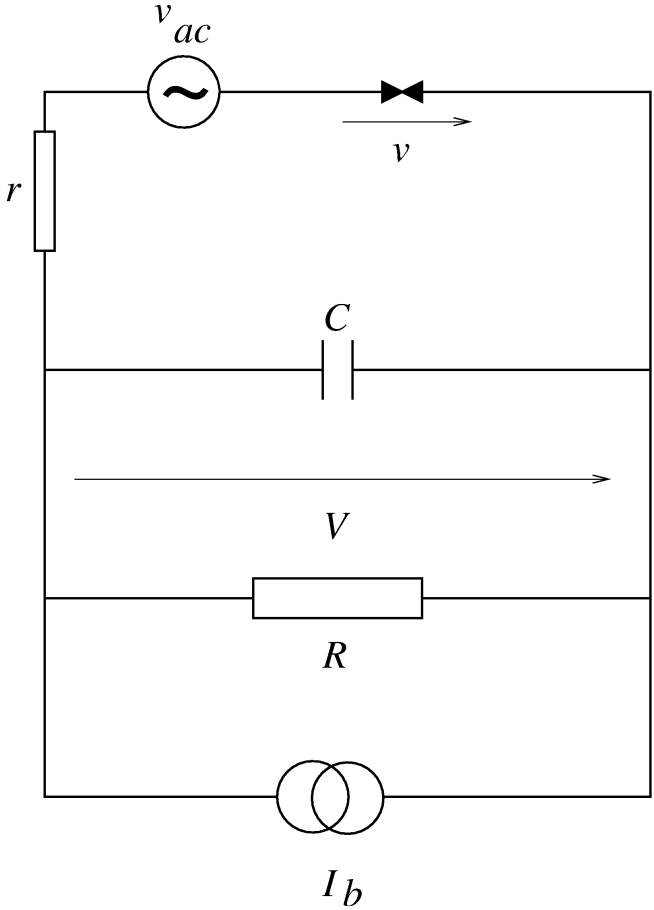}
\caption{\label{fig:circuit2R} Generalization of the RCSJ model.}
\end{figure}

Neglecting the capacitance in the usual RCSJ model leads to considerable
simplification of the circuit equations. However this may result in a
rather crude description of the electromagnetic environment in an actual
experiment.  We consider in this section a more realistic circuit with
two resistors and a capacitance as shown in Fig. \ref{fig:circuit2R}. 
This model reduces to the previous one 
when $r = 0$. The presence of two resistors means that there are now two 
sources of thermal noise. We thus have to deal with two coupled Langevin 
equations:
\begin{eqnarray}
C\frac{dV}{dt} &=& I_b - I(\theta) - \frac{V}{R} - L_R(t) \\
\phi_0\frac{d\theta}{dt} &=& V - rI(\theta) - v_{ac}\cos\omega t - rL_r(t) \ .
\end{eqnarray}
Using standard techniques (see Ref. \onlinecite{risken} for instance), we can 
derive from them a Fokker-Planck equation for the probability density
$W(\theta,V,t)$:
\begin{widetext}
\begin{equation} \label{FPE}
\frac{\partial W}{\partial t} = -\frac{\partial}{\partial\theta}\Bigl[\frac{V-v_{ac}\cos\omega t -rI(\theta)}{\phi_0} W \Bigr] -\frac{\partial}{\partial V} \Bigl[ \frac{RI_b - RI(\theta) - V}{RC} W \Bigr] + \frac{T}{RC^2}\frac{\partial^2 W}{\partial \theta^2} + \frac{rT}{\phi_0^2}\frac{\partial^2 W}{\partial V^2} \ .
\end{equation}
\end{widetext}
Setting $r=0$ and noting that in the strong damping limit 
$\frac{\partial}{\partial V} = \frac{1}{\eta\phi_0}
\frac{\partial}{\partial\theta}$, one easily recovers the Smoluchowski 
equation (\ref{smolu}).

In order to solve the Fokker-Planck equation (\ref{FPE}), we need to
find on which basis to expand the $V$ dependent part of the distribution 
function
$W$. Following Ref. \onlinecite{risken}, we split the Fokker-Planck
operator in a reversible and an irreversible part, and note that the
$V$-dependent term of the latter coincides with the Hamiltonian of a harmonic 
oscillator when brought to an hermitian form. This suggests to consider Hermite polynomials $H_k(z)$. More precisely, we will use the following orthonormal 
basis,
\begin{equation}
\psi_{k\in\mathbb{N}}(V) = e^{\frac{-CV^2}{4T}} H_k\Bigl(V\sqrt{\frac{C}{2T}}\Bigr) [k!2^k\sqrt{2\pi MT}]^{-\frac{1}{2}} \ .
\end{equation}
The distribution will then read
\begin{equation}
W(\theta,V,t)=\psi_0(V)\sum_{n m k}\psi_k(V)e^{in\omega t +im\theta} W_{n m k} \ ,
\end{equation}
and the normalization implies
\begin{equation}
W_{n 0 0} = \frac{\delta_{n,0}}{2\pi} \ .
\end{equation}
The expressions of the mean voltage across the resistance $R$ and of the mean current and voltage across the contact are then
\begin{eqnarray}
<<\bar{V}>> &=&  \sqrt{\frac{T}{C}} W_{001} \\
<<\overline{I(\theta)}>> &=&  \sum_m I_m W_{0,-m,0} \\
<<\bar{v}>> &=& <<\bar{V}>>  - r\sum_m I_m W_{0,-m,0} \ ,
\end{eqnarray}
where $<<\overline{(...)}>> \equiv \int\int\int dt dV d\theta (...)$ represents the mean value over the time, the voltage and the phase. Thus, again, the knowledge of the $n=0$ component of the distribution is sufficient to 
evaluate the quantities of interest.

The irreversible part of the Fokker-Planck operator being diagonal in the basis $\{\psi_k(z) ; k\in\mathbb{N}\}$, we can anticipate a simple recurrence relation between the components of the distribution in the voltage sub-space, which should lead to a continued fraction of matrix continued fractions generalizing 
(\ref{MCF1}).

Making use of the orthogonality of circular and $\psi_k$ functions, together with the recurrence relation for Hermite polynomials,
\begin{equation}
H_k(z) = 2zH_{k-1}(z)-2(k-1)H_{k-2}(z) \ ,
\end{equation}
we can rewrite the Fokker-Planck equation as
\begin{widetext}
\begin{eqnarray} \label{Hermite-Fourier}
in\hbar\omega W_{n,m,k} &=& A^{k,k}_{m,m'} W_{n,m',k} + A^{k,k+1}_{m,m}
W_{n,m,k+1} + A^{k,k-1}_{m,m'} W_{n,m',k-1} \nonumber \\
& & + i mev_{ac}\left(W_{n+1,m,k}
+ W_{n-1,m,k} \right)
\end{eqnarray}
\end{widetext}
where
\begin{eqnarray}
A^{k,k}_{m,m'} & = & -\delta_{m,m'} \left( \frac{2k}{\pi} \frac{R_Q}{R} 
\frac{e^2}{C} + 2 \pi \frac{r}{R_Q} T m^2 \right) + \nonumber \\
&& (1-\delta_{m,m'}) 2\pi i m \frac{r}{R_Q} \phi_0 I_{m-m'}\\
A^{k,k-1}_{m,m'} &=& \delta_{m,m'} \left(2 \phi_0 I_b \sqrt{\frac{e^2}{CT}k}
- 2 i m \sqrt{\frac{e^2}{C} T k} \right) - \nonumber \\
&& (1-\delta_{m,m'}) 2 \phi_0 I_{m-m'} \sqrt{\frac{e^2}{CT}k} \\
A^{k,k+1}_{m,m} &=& 2 i m \sqrt{\frac{e^2}{C} T (k+1)}. 
\end{eqnarray}

\begin{figure}
\includegraphics[width=\columnwidth]{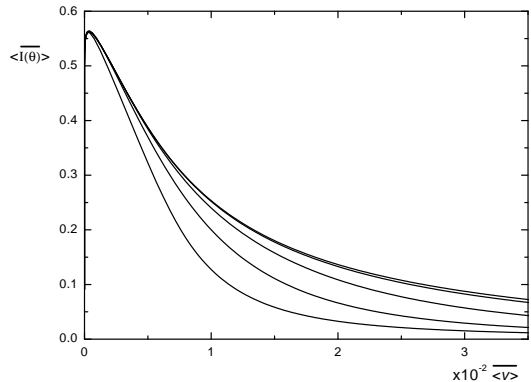}
\caption{\label{fig:GFP} Supercurrent peak for a contact with 
$\tau=0.9$ in the circuit of Fig. \ref{fig:circuit2R} with 
$R=0.001$, $r=R/10$, $T=0.02$ and $e^2/C=0.0001$, 0.0002, 0.0004, 0.0008
and 0.0016 from bottom to top. As can be observed the results converge
to an asympthotic curve corresponding to the RSJ model.}
\end{figure}

Exploting its block-tridiagonal structure the set of equations
(\ref{Hermite-Fourier}) can be evaluated using a recursive algorithm
similar to the one discussed in Sect. III. In order to illustrate the
effect of this more complex environment we concentrate in the analysis 
of the supercurrent peak in the absence of microwaves \cite{comment-GFP}. 
Figure \ref{fig:GFP} shows the current-voltage characteristic   
around zero bias for a contact with $\tau=0.9$,  
$r=R/10$ and different values of the capacitance in the circuit. 
As can be observed the width of the supercurrent peak tends to increase
as the size of the capacitance is reduced. At the same time the height of 
the peak exhibits a slight reduction, which is almost negligible at low
temperatures. The shape of the supercurrent peak approaches the 
one found in the standard RSJ model when $e^2/C \ge 2\pi R$.

\section{Conclusion}
We have studied the effect of classical phase diffusion 
on fractional Shapiro steps 
in quantum point contacts. For this  
purpose we have generalized the standard 
RCSJ model to superconducting contacts with arbitrary transmission in the 
presence of a microwave field. In the overdamped limit the circuit equations 
can be mapped into a Smoluchowski equation for the probability density of 
the phase difference across the contact. We have presented an efficient 
algorithm for the numerical evaluation of this equation.
It has been shown that the fractional steps exhibit a stronger suppression 
with temperature than the interger ones in agreement with preliminary 
experimental findings \cite{saclay}. In the limit of large microwave 
amplitude Shapiro steps of order $n/k$ exhibit an approximate universal 
behavior as a function of an effective temperature $T_{eff} = k^2T$ and
an effective microwave parameter $\alpha_{eff} = k \alpha$.
We have also considered the case of a more realistic environment 
including two resistances and a finite capacitance for which we derived the 
corresponding Fokker-Plank equation. Our numerical results indicate that the 
main effect of the finite capacitance is to reduce the width of the 
supercurrent peak. We expect that the theoretical analysis presented in this 
work may be useful for the proper interpretation of future experiments.

\begin{acknowledgments}
The authors would like to thank J. Ankerhold, M. Chauvin, J.C. Cuevas, 
D. Est\`eve, P. Joyez, A. Mart\'in-Rodero, P. vom Stein and C. Urbina for 
interesting discussions and suggestions. This work was supported by the 
European Union through the 
DIENOW Research Training Network and by the Spanish CICyT through 
a Picasso contract.
\end{acknowledgments}


\begin{thebibliography}{}


\bibitem{shapiro63}
S. Shapiro, Phys.\ Rev.\ Lett. {\bf 11}, 80 (1963). 

\bibitem{josephson62}
B. D. Josephson, Phys.\ Lett {\bf 1}, 251 (1962).

\bibitem{cuevas02}
J. C. Cuevas, J. Heurich, A. Mart\'in-Rodero, A. {Levy Yeyati} 
and G. Sch{\"o}n, Phys.\ Rev.\ Lett. {\bf 88}, 157001 (2002). 

\bibitem{review03}
For a review see, N. {Agra\"{\i}t}, A. {Levy Yeyati} and J.M. van Ruitenbeek,
Phys. Rep. {\bf 377}, 81 (2003).

\bibitem{scheer97}
E. Scheer, P. Joyez, D. Esteve, C. Urbina and M.H.
Devoret, Phys. Rev. Lett., {\bf 78}, 3535 (1997);
E. Scheer, N. {Agra\"{\i}t}, J.C. Cuevas, A. {Levy
Yeyati}, B. Ludoph, A. Mart{\'\i}n-Rodero, G. {Rubio Bollinger},
J.M. van Ruitenbeek and C. Urbina, Nature {\bf 394}, 154 (1998).

\bibitem{goffman00}
M. F. Goffman, R. Cron, A. {Levy Yeyati}, P. Joyez, M. H. Devoret,
D. Esteve and C. Urbina, Phys.\ Rev.\ Lett.{\bf 85}, 170 (2000). 

\bibitem{tinkham}
M. Tinkham,
{\it Introduction to superconductivity}
(McGraw-Hill, New-York, 1996).

\bibitem{ivan69}
Yu. M. Ivanchenko and L. A. Zil'berman, Sov.\ Phys.-JETP {\bf 28}, 1272 (1969).

\bibitem{ambegaokar69}
V. Ambegaokar and B. I. Halperin,
Phys.\ Rev.\ Lett. {\bf 22}, 1364 (1969).

\bibitem{saclay}
P. Vom Stein, M. Chauvin, R. Duprat, A. Levy
Yeyati, M. E.\ Huber, D. Esteve, and C. Urbina,
to be published in {\it Proceedings of the Vth Rencontres de Moriond in
Mesoscopic Physics}. 

\bibitem{vanneste} P. Russer, J. Appl. Phys. {\bf 43}, 320 (1972);
C. Vanneste et al, Phys. Rev. B {\bf 31}, 4230 (1985).

\bibitem{schwabl}
F. Schwabl, {\it Statistical mechanics},
(Springer, Berlin, 2002).

\bibitem{beenakker92}
C.W.J. Beenakker in
{\it Proceedings of the 14th Taniguchi International Symposium
on Transport Phenomena in Mesoscopic Systems},
edited by H. Fukuyama and T. Ando
(Springer-Verlag, Berlin, 1992), p. 235.

\bibitem{ambegaokar63}
V. Ambegaokar and A. Baratoff,
Phys.\ Rev.\ Lett. {\bf 10},
486, (1963).

\bibitem{carvello02}
B. Carvello and C. Urbina, unpublished.

\bibitem{nazarov}
See for instance G.-L. Ingold and Yu.V. Nazarov in {\it Single Charge
Tunneling}, edited by H. Grabert and M.N. Devoret (Plenum Press, New York,
1992).


\bibitem{risken}
H. Risken,
{\it The Fokker-Planck equation}
(Springer-Verlag, Berlin 1989).

\bibitem{comment-GFP}
For the Shapiro steps we expect a qualitatively similar effect
due to the finite capacitance as for the supercurrent peak.

\end{thebibliography}
\end{document}